\documentclass[12pt,a4paper,epsf]{article}
\usepackage{graphics}
\usepackage{amssymb,amsmath}
\usepackage[dvips]{lscape,graphicx}
\usepackage{cite}

\newcommand{\ct}{\cite}
\newcommand{\lb}{\label}

\newcommand{\bc}{\begin{center}}
\newcommand{\ec}{\end{center}}
\newcommand{\bd}{\begin{displaymath}}
\newcommand{\ed}{\end{displaymath}}
\newcommand{\be}{\begin{equation}}
\newcommand{\ee}{\end{equation}}
\newcommand{\ba}{\begin{array}}
\newcommand{\ea}{\end{array}}
\newcommand{\bea}{\begin{eqnarray}}
\newcommand{\eea}{\end{eqnarray}}
\newcommand{\bt}{\begin{tabular}}
\newcommand{\et}{\end{tabular}}

\newcommand{\bp}{\begin{picture}}
\newcommand{\ep}{\end{picture}}
\newcommand{\bfi}{\begin{figure}}
\newcommand{\efi}{\end{figure}}

\begin{document}

\hyphenation{ }

\title{\Large\bf Phase Transition in the Higgs Model of Scalar Dyons}

\author{\large
C.R.~Das ${}^{1}$ \footnote{\large\, crdas@imsc.res.in} ,
L.V.~Laperashvili ${}^{1,\, 2}$ \footnote{\large\, laper@itep.ru, laper@imsc.res.in}\\[5mm]
\itshape{${}^{1}$ The Institute of Mathematical Sciences, Chennai, India}
\\[0mm]
\itshape{${}^{2}$ The Institute of Theoretical and Experimental Physics, Moscow, Russia}}

\date{}
\maketitle

\thispagestyle{empty}

\vspace{1cm}

\begin{abstract}

In the present paper we investigate the phase transition ``Coulomb--confinement" 
in the Higgs model of abelian scalar dyons -- 
particles having both, electric $e$ and magnetic $g$, charges. It is shown that by
dual symmetry this theory is equivalent to scalar fields with the effective squared electric 
charge $e^{*2}=e^2+g^2$. But the Dirac relation distinguishes the electric and magnetic charges 
of dyons. The following phase transition couplings are obtained in the one--loop approximation:
$\alpha_{crit}=e^2_{crit}/4\pi\approx 0.19$, $\tilde\alpha_{crit}=g^2_{crit}/4\pi\approx 
1.29$ and $\alpha^*_{crit}\approx 1.48$.

\end{abstract}

\clearpage\newpage

\setcounter{page}{1}

\noindent{\bf 1.}
In the present paper we consider dyons -- the particles with electric $e$ and magnetic $g$ 
charges. As it was shown in Refs.~\ct{1,1aa,2,3,4,5}, dyons play an essential role in 
physics of nonabelian theories (in particular, in QCD).

The local field theory of electrically and magnetically charged particles, so called 
``QuantumElectroMagnetoDynamics'' (QEMD) \ct{6}, is presented by the Zwanziger 
formalism \ct{7,8}, (see also \ct{9} and \ct{10}), which considers two vector potentials 
$A_\mu(x)$ and $B_\mu(x)$, describing one physical photon with two physical degrees of freedom. 
Here $B_{\mu}$ is the magnetic gauge potential, which is dual to the electric gauge 
potential $A_{\mu}$. This formalism symmetrically contains
non--dual and dual interactions of gauge fields with the corresponding currents: $j^{(e)}\cdot
A$ and  $j^{(m)}\cdot B$, where $j_\mu^e$ and $j_\mu^m$ are electric and magnetic currents, 
respectively.

Dyons are described by the field $\Phi$, having charges $n_1e$ and $n_2g$ $(n_1, n_2 \in Z)$.
The total system of gauge fields and dyons is given by the partition function having the
following form in Euclidean space:
\be Z=\int [DA][DB][D\Phi][D\Phi^+]e^{-S}, \lb{1} \ee
where
\be S = \int d^4x\,L(x)=S_{Zw}(A,B)+S_{gf}+S_{(matter)}. \lb{2} \ee
The Zwanziger action $S_{Zw}(A,B)$ is given by:
\bea S_{Zw}(A,B) &=& \int d^4x \left[\frac 12 {(n\cdot[\partial \wedge A])}^2 +
\frac 12 {(n\cdot[\partial \wedge B])}^2 +\right.\nonumber\\
&&\left.+\frac i2(n\cdot[\partial \wedge A])(n\cdot{[\partial \wedge B]}^*)
- \frac i2(n\cdot[\partial \wedge B])(n\cdot{[\partial \wedge A]}^*)\right],\lb{3}\eea
where we have used the following designations:
$${[A \wedge B]}_{\mu\nu} = A_{\mu}B_{\nu} - A_{\nu}B_{\mu},
 \qquad {(n\cdot[A \wedge B])}_{\mu} = n_{\nu}{(A \wedge B)}_{\nu\mu}, $$
\be {G}^*_{\mu\nu} = \frac 12\epsilon_{\mu\nu\lambda\rho}G_{\lambda\rho}. \lb{4}\ee
In Eqs.~(\ref{3}) and (\ref{4}) the unit vector $n_{\mu}$ represents
the fixed direction of the Dirac string in the 4--space.

The action $S_{(matter)}$:
\be S_{(matter)} = \int d^4x\, L_{(matter)}(x)\lb{5}\ee
describes the matter fields of dyons.

In Eq. (\ref{2}) $S_{gf}$ is the gauge--fixing action. According to Ref.~\ct{9}, this action is 
given by the following equation:
\be S_{gf} = \int d^4x \left[\frac{M^2_A}2{(n\cdot A)}^2 + \frac{M^2_B}2{(n\cdot B)}^2\right],
\lb{6}\ee
and has no ghosts. Eq.~(\ref{6}) contains the mass parameters $M_A$ and $M_B$.

Let us consider now the Lagrangian $L_{(matter)}$, containing the Higgs scalar dyon field  
$\Phi(x)$ interacting with gauge fields $A_{\mu}(x)$ and $B_{\mu}(x)$:
\be 
L_{(matter)}(x) = \frac 12 {|{\cal D}_{\mu}\Phi|}^2  - U(\Phi), \lb{7} 
\ee
where
\be {\cal  D}_{\mu} = \partial_{\mu} - ieA_{\mu} - igB_{\mu}, 
\ee
and 
\be 
    U(\Phi) = \frac 12 m^2{|\Phi|}^2 + \frac{\lambda}4{|\Phi|}^4  \lb{9}
\ee
is the Higgs potential for the dyon field $\Phi$.

The complex scalar fields:
\be \Phi = \phi + i\chi \lb{10}\ee
contain the Higgs $(\phi )$ and Goldstone $(\chi)$ boson fields.

In the Lagrangian (\ref{7}) the interactions, given by terms $j_{\mu}^eA_{\mu}$ and 
$j_{\mu}^mB_{\mu}$, contain the following  electric and magnetic currents:
\bea j_{\mu}^e &=& ie[\Phi^+{\cal D}_{\mu}\Phi - \Phi({\cal D}_{\mu}\Phi)^+], \lb{11}\\
 j_{\mu}^m &=& ig[\Phi^+{\cal\tilde D}_{\mu}\Phi 
- \Phi({\cal\tilde D}_{\mu}\Phi)^+].\lb{12}\eea
In Eq.~(\ref{7}) we have ``sea--gull" terms $e^2A_{\mu}^2{|\Phi|}^2$ and 
$g^2B_{\mu}^2{|\Phi|}^2$.

Letting
\bea F &=& \partial\wedge A = - (\partial\wedge B)^{*},\lb{13}\\
F^{*} &=& \partial\wedge B = (\partial\wedge A)^{*},\lb{14}\eea
and the Hodge star operation (\ref{4}) on the field tensor:
\be F_{\mu\nu}^{*} = \frac 12 \epsilon_{\mu\nu\rho\sigma} F_{\rho\sigma}, \lb{15}\ee
it is easy to see that the free Zwanziger Lagrangian (\ref{3})
is invariant under the following duality transformations:
\be F \leftrightarrow F^{*}, \qquad (\partial\wedge A) \leftrightarrow (\partial \wedge B),
 \qquad (\partial\wedge A)^{*} \leftrightarrow - (\partial \wedge B)^{*}. \lb{16}\ee
We also have a dual symmetry as an invariance of the total
Lagrangian $L(x)$, provided that electric and magnetic charges and currents
transform simultaneously according to the following discrete symmetry:
\bea &&e \rightarrow g, \qquad g \rightarrow - e,\nonumber\\
&&j_{\mu}^e \rightarrow j_{\mu}^m, \qquad j_{\mu}^m \rightarrow - j_{\mu}^e.\lb{17}\eea
The action (\ref{2}), given by Eqs.~(\ref{3}) and (\ref{5}-\ref{9}), leads to the following 
field equations:
\be \partial_{\lambda} F_{\lambda \mu} = j_{\mu}^e, \qquad \partial_{\lambda} 
F_{\lambda \mu}^{*} =  j_{\mu}^m, \lb{18}\ee 
where currents $j_{\mu}^e$, $j_{\mu}^m$ are given by Eqs.~(\ref{11}, \ref{12}).  

\vspace{1cm}\noindent{\bf 2.}
The Lorentz invariance is lost in the Zwanziger Lagrangian (\ref{3}), because it depends on a 
fixed vector $n_{\mu}$. However, this invariance is regained for the quantized values of 
coupling constants $e$ and $g$, obeying the Dirac--Schwinger--Zwanziger (DSZ) relation for 
dyons:
\be e_ig_j -  g_ie_j = 2\pi n_{ij}, \qquad n_{ij} \in Z. \lb{19}\ee
As it was shown in Refs.~\ct{11,11a,11b,12,12a,13} (see also \ct{14a}), gauge theories 
(abelian and nonabelian) can be constructed in terms of loop variables.
The new Zwanziger--type Lagrangian does not contain a fixed vector $n_{\mu}$, but
contains loop coordinates and their derivatives.
Considering the closed loop operators:
\be A(C) = P\exp \left[ie\oint_C A_{\mu}(\xi) d{\xi}^{\mu}\right], \lb{20}\ee 
which measures magnetic flux through $C$ and creates electric flux along $C$, and: 
\be B(\tilde C) = P \exp \left[ ig\oint_{\tilde C} {\tilde A}_{\mu}(\eta)d{\eta}^{\mu}\right], 
\lb{21} \ee 
which  measures electric flux through $\tilde C$ and creates magnetic flux along $\tilde C$, 
we have the t'Hooft commutation relation \ct{14}:
\be A(C)B(\tilde C) = B(\tilde C)A(C)\exp(2\pi i n), \lb{22}\ee 
where $n$ is the number of times $\tilde C$ winds around $C$. In Eqs.~(\ref{20}) and (\ref{21}) 
we have coordinates $\xi^{\mu}$ and $\eta^{\mu}$ of the loops $C$ and $\tilde C$ in the 
4--dimensional space.

Eq.~(\ref{22}) gives the Dirac relation -- the charge quantization condition for a single dyon,
interacting with fields $A_{\mu}$ and $B_{\mu}$:
\be eg = 2\pi n, \qquad n\in Z.  \lb{23}\ee
For elementary particles we have:
\be eg = 2\pi, \qquad \mbox{or} \qquad \alpha\tilde\alpha=\frac 1 4,\lb{24}\ee
where $\alpha$ and $\tilde\alpha$ are the electric and magnetic fine structure constants:
\be\alpha=\frac{e^2}{4\pi} \qquad \mbox{and} \qquad \tilde\alpha=\frac{g^2}{4\pi}.\lb{25}\ee

\vspace{1cm}\noindent{\bf 3.}
The effective potential in the Higgs model of scalar electrodynamics was calculated in the 
one-loop approximation for the first time by authors of Ref.~\ct{15}. A general method of
calculation of the effective potential is given in the review \ct{16}. Using this method, we 
can construct the effective potential for theory described by the partition function (\ref{1}) 
with the action $S$, containing the Zwanziger action (\ref{3}), gauge fixing action (\ref{6}) 
and the action (\ref{7}) for dyon matter fields.

Let us consider now the shift:
\be\Phi (x) = \Phi_b + {\hat \Phi}(x) \lb{26}\ee
with $\Phi_b$ as a background field, and calculate the following expression for the partition 
function in the one-loop approximation:
\bea Z &=& \int [DA][DB][D\hat \Phi][D{\hat \Phi}^{+}] \nonumber\\
&&\cdot \exp\left\{ - S(A,B,\Phi_B) - \int d^4x 
\left[\left.\frac{\delta S(\Phi)}{\delta \Phi(x)}\right|_{\Phi=
\Phi_B}{\hat \Phi}(x) + h.c. \right]\right\}\nonumber\\
&=&\exp\{ - F(\Phi_b, e^2, g^2, m^2, \lambda)\}.\lb{27}\eea
Using the representation (\ref{10}), we obtain the effective potential:
\be V_{eff} = F(\phi_b, e^2, g^2, m^2, \lambda), \lb{28}\ee
given by the function $F$ of Eq.~(\ref{27}) for the constant background field:
\be\Phi_b = \phi_b = \mbox{const}. \lb{29}\ee
The effective potential (\ref{28}) has several minima. Their position depends on $e^2$, $g^2$, 
$m^2$ and $\lambda$. If the first local minimum occurs at $\phi_b=0$, it corresponds to the 
so--called ``symmetrical phase", which is the ``Coulomb" phase in our description.

We are interested in the phase transition from the Coulomb phase ``$\phi_b = 0$" to the 
confinement phase ``$\phi_b = \phi_0 \neq 0$". In this case the one--loop effective potential 
for dyons is similar to the expression of the effective potential, calculated by authors of 
Ref.~\ct{15} for scalar electrodynamics and extended to the massive theory in Ref.~\ct{17}.

As it was shown in Ref.~\ct{15}, effective potential can be improved by the consideration of 
the renormalization group equation (see also the review \ct{16}).

\vspace{1cm}\noindent{\bf 4.}
The renormalization group (RG) describes an independence of a theory and its couplings on an 
arbitrary scale parameter $M$. We are interested in RG applied to the effective potential. The 
renormalization group equation (RGE) for the effective potential means that the potential 
cannot depend on a change in the arbitrary renormalization scale parameter $M$:
\be\frac{dV_{eff}}{dM}.\lb{30}\ee
The effects of changing it are absorbed into changes in the coupling constants, masses and 
fields, giving so--called running quantities. Knowing the dependence on $M^2$ is equivalent to 
knowing the dependence on $\phi^2\equiv\phi_b^2$. This dependence is given by RGE. Considering 
the RGE improvement of the potential, we follow the approach by Coleman and Weinberg \ct{15} 
and its extension to the massive theory \ct{17}. Here we have the difference between the scalar 
electrodynamics \ct{15} and scalar QEMD.

RGE for the improved one--loop effective potential can be given in QEMD by the following 
expression:
\be\left(M\frac{\partial}{\partial M} +\beta_{\lambda}\frac{\partial}{\partial \lambda} +
e\beta_e\frac{\partial}{\partial e} + g\beta_g\frac{\partial}{\partial g} +
\beta_{(m^2)}{m^2}\frac{\partial}{\partial m^2} - \gamma \phi 
\frac{\partial}{\partial \phi}\right) V_{eff}(\phi^2) = 0, \lb{31}\ee
where the function $\gamma $ is the anomalous dimension:
\be \gamma\left(\frac{\phi}M\right) = - \frac{\partial \phi}{\partial M}. \lb{32}\ee
RGE (\ref{31}) leads to a new improved effective potential \ct{15}:
\be V_{eff}(\phi^2) = \frac 12 m^2_{ren}(t) G^2(t)\phi^2 +
\frac 14 \lambda_{ren}(t) G^4(t) \phi^4, \lb{33}\ee
where
\be G(t)\equiv \exp\left[ - \frac 12 \int_0^t dt' \gamma(g_{ren}(t'),
\lambda_{ren}(t'))\right]. \lb{34}\ee
Eq.~(\ref{31}) reproduces also a set of ordinary differential equations:
\be\frac{d\lambda_{ren}}{dt} = \beta_{\lambda}(g_{ren}(t),\lambda_{ren}(t)), \lb{35}\ee
\be\frac{dm^2_{ren}}{dt} = m^2_{ren}(t) \beta_{(m^2)}(g_{ren}(t),\lambda_{ren}(t)), \lb{36}\ee
\be\frac{d\ln e_{ren}}{dt} = - \frac{d\ln g_{ren}}{dt} = \beta_e(g_{ren}(t),
\lambda_{ren}(t)) - \beta_g (g_{ren}(t),\lambda_{ren}(t))\equiv \beta^{(total)}, \lb{37}\ee
where $t = \ln (\phi^2/{M^2})$, and the subscript ``ren" means the ``renormalized" quantity.

The last equation (\ref{37}) is obtained with the help of the Dirac relation (\ref{24}) for 
minimal charges. Indeed, in Eq.~(\ref{31}):
$$ e\beta_e\frac{\partial}{\partial e} + g\beta_g\frac{\partial}{\partial g} =  
e\beta_e\frac{\partial}{\partial e} + g\beta_g\frac{de}{dg}\frac{\partial}{\partial e} 
= \left(e\beta_e + g\beta_g\left(-\frac{2\pi}{g^2}\right)\right)\frac{\partial}{\partial e} $$ 
\be = (\beta_e - \beta_g)\frac{\partial}{\partial \ln e} 
   = \beta^{(total)}\frac{\partial}{\partial \ln e},   \lb{38} \ee
where
\be\beta^{(total)} = \beta_e - \beta_g.  \lb{39}\ee
We can determine both beta functions for $\lambda_{ren}$ and $m^2_{ren}$ by considering a 
change in $M$ in the conventional non--improved one--loop potential.

Let us write now the one--loop potential: 
\be V_{eff} = V_0 + V_1, \lb{40}\ee
where
\be V_0 = \frac{m^2}2 \phi^2 + \frac{\lambda}4 \phi^4, \lb{41}\ee
and (see Refs.~\ct{15} and \ct{16}):
\bea V_1 &=& \frac{1}{64\pi^2}\left[ 3(e^2 + g^2)^2 {\phi}^4\ln\frac{\phi^2}{M^2}
+ {(m^2 + 3\lambda {\phi}^2)}^2\ln\frac{m^2 + 3\lambda\phi^2}{M^2}\right.\nonumber\\
   &&\left.+ {(m^2 +\lambda \phi^2)}^2\ln\frac{m^2
+ \lambda \phi^2}{M^2} - 2m^4\ln \frac{m^2}{M^2}\right]. \lb{42}\eea
We can plug $V_{eff}$, given by Eqs.~(\ref{40}-\ref{42}), into RGE (\ref{31}) and obtain the
following equation (see \ct{16}):
\be\left( \beta_{\lambda}\frac{\partial}{\partial \lambda} +
 \beta_{(m^2)}{m^2}\frac{\partial}{\partial m^2} -
 \gamma \phi \frac{\partial}{\partial \phi}\right) V_0 =
 - M\frac{\partial V_1}{\partial M}. \lb{43}\ee
Equating $\phi^2$ and $\phi^4$ coefficients, we obtain:
\be\beta_{\lambda} = 2\gamma \lambda_{ren} + \frac{5\lambda_{ren}^2}{8\pi^2} +
\frac{3(e_{ren}^2 + g_{ren}^2)^2}{16\pi^2}, \lb{44}\ee
\be \beta_{(m^2)} = \gamma + \frac{\lambda_{ren}}{4\pi^2}. \lb{45}\ee
The result for $\gamma$ is given in Ref.~\ct{15} for scalar field with electric charge $e$, 
but it is easy to rewrite this $\gamma$--expression for dyons with renormalized charges 
$e_{ren}$ and $g_{ren}$:
\be\gamma = - \frac{3(e_{ren}^2 + g_{ren}^2)}{16\pi^2}. \lb{46}\ee
Finally we have:
\be\frac{d\lambda_{ren}}{dt} = \frac 1{16\pi^2}\left[ 3(e_{ren}^2 + g^2_{ren})^2 +10 
\lambda^2_{ren} - 6\lambda_{ren}(e^2_{ren} + g^2_{ren})\right], \lb{47}\ee
\be\frac{dm^2_{ren}}{dt} = \frac{m^2_{ren}}{16\pi^2}\left[4\lambda_{ren}^2 -
                           3(e^2_{ren} + g^2_{ren})\right]. \lb{48}\ee

Now the aim is to calculate the beta--function $\beta^{(total)}$
in Eq.~(\ref{37}).

\vspace{1cm}\noindent{\bf 5.}
Many years ago it was calculated (see Refs.~\ct{18,18a,19,19a,19b}) that 
Gell--Mann--Low equation \ct{20}:
\be \frac{d\ln \alpha}{dt} = \beta(\alpha) \lb{49}\ee
has the following $\beta$-function for the electric charge in the scalar electrodynamics:
\be\beta(\alpha) = \frac{\alpha}{12\pi}\left(1+3\frac{\alpha}{4\pi} + ...\right). \lb{50}\ee
According to the previous item 4, now we have:
\be\frac{ d\ln \alpha}{dt} = -\frac{ d\ln {\tilde \alpha}}{dt} =
\beta (\alpha ) - \beta(\tilde \alpha ) = \beta^{(total)}(\alpha). \lb{51}\ee
These RGEs are a consequence of the Dirac relation (\ref{24}) and the dual symmetry, 
considered in the item 1. 

If both $\alpha$ and $\tilde{\alpha}$ are sufficiently small, then beta--functions in 
Eq.~(\ref{51}) are described by the contributions of the electrically and magnetically charged 
dyon loops simultaneously. Their analytical expressions are given by (\ref{50}), and in the 
two--loop approximation we have the following equations (\ref{51}) for scalar dyons 
(see also \ct{10}):
\be\frac{ d\ln \alpha}{dt} = - \frac{ d\ln {\tilde \alpha}}{dt} =
\frac{\alpha - \tilde \alpha }{12\pi}\left( 1 + 3\frac{\alpha
+ \tilde \alpha}{4\pi} + ....\right)= \beta^{(total)}(\alpha). \lb{52}\ee
According to Eq.~(\ref{52}), the two--loop contribution is not more than $30\%$ if both 
$\alpha$ and $\tilde {\alpha}$ obey the following requirement (see Ref.~\ct{10}):
\be 0.25 \lesssim \alpha, \tilde{\alpha}\lesssim 1. \lb{53}\ee
The lattice simulations of compact QED give the behaviour of the effective fine structure
constant in the vicinity of the phase transition point \ct{21,22,23,24,25}. The following 
critical values of the fine structure constants $\alpha$ and $\tilde\alpha$ were obtained in 
Ref.~\ct{21}:
\be\alpha_{crit}^{lat}\approx{0.20 \pm 0.015}, \qquad 
{\tilde \alpha}_{crit}^{lat}\approx{1.25 \pm 0.10}. \lb{54}\ee
Eq.~(\ref{54}) demonstrates that $\alpha_{crit}$ and $\tilde\alpha_{crit}$, obtained in the
compact lattice QED \ct{21,22,23,24,25}, almost coincide with the borders of the requirement 
(\ref{53}), given by the perturbation theory for $\beta$--function \ct{10}. Assuming that in 
the vicinity of the phase transition point the coupling constant $g_{ren}$ may be described by 
the one--loop approximation, we obtain from Eq.~(\ref{52}) the following RGE for $g_{ren}^2$:
\be\frac{d\ln(g_{ren}^2)}{dt} = \frac{g^2_{ren} - e^2_{ren}}{48\pi^2}. \lb{55}\ee
Using the Dirac relation (\ref{24}), we have:
\be\frac{dg^2_{ren}}{dt} = \frac{g^4_{ren}}{48\pi^2} - \frac 1{12}. \lb{56}\ee
Note that the second term of Eq.~(\ref{56}) describes the influence of the electric charge on 
the behaviour of the magnetic one.

\vspace{1cm}\noindent{\bf 6.}
In this part of our paper we investigate the phase transition from the ``Coulomb'' phase 
($\phi_b = 0$) to the confinement one ($\phi_b= \phi_0 \neq 0$), following the methods of 
Refs.~\ct{26,26a,27}. This means that the effective potential (\ref{33}) of the Higgs scalar 
dyons 
has the first and the second minima appearing at $\phi = 0$ and $\phi = \phi_0$, respectively. 
They are shown in Fig.~\ref{f1} by the solid curve ``1". These minima of $V_{eff}(\phi^2)$ 
correspond to the different vacua arising in this model. The conditions for the existence of 
degenerate vacua are given by the following equations:
\be V_{eff}(0) = V_{eff}(\phi_0^2) = 0, \lb{57}\ee
\be\left.\frac{\partial V_{eff}}{\partial \phi}\right|_{\phi=0} =
\left.\frac{\partial V_{eff}}{\partial \phi}\right|_{\phi=\phi_0} = 0\lb{58}\ee
with inequalities
\be \left.\frac{\partial^2 V_{eff}}{\partial \phi^2}\right|_{\phi=0} > 0, \qquad 
\left.\frac{\partial^2 V_{eff}}{\partial \phi^2}\right|_{\phi=\phi_0} > 0,\lb{59}\ee 
or considering $\phi^2$ as a variable, we can write:
\bea V'_{eff}(\phi_0^2) &\equiv&\left. \frac{\partial V_{eff}}{\partial \phi^2}\right|_
{\phi=\phi_0} = 0, \lb{60}\\
V''_{eff}(\phi_0^2)&\equiv&\left.\frac{\partial^2 V_{eff}}{\partial {(\phi^2)}^2}\right|_
{\phi=\phi_0} > 0. \lb{61}\eea
From now we omit (for simplicity) the subscript ``ren", using the following designations: 
$$e, g, \lambda, m^2\equiv e_{ren},\, g_{ren},\, \lambda_{ren},\, m^2_{ren}.$$ 
The first equation (\ref{57}) applied to Eq.~(\ref{33}) gives:
\be m^2 = - \frac 12 \lambda(t_0){\phi_0^2}G^2(t_0), \qquad 
\mbox{where} \qquad t_0 = \ln \left(\frac{\phi_0^2}{M^2}\right)\lb{62}.\ee

The calculation of the first derivative of $V_{eff}(\phi^2)$ leads to the following expression:
\bea V'_{eff}(\phi^2) &=& \frac{V_{eff}(\phi^2)}{\phi^2}\left(1 + 2\frac{d\ln G}{dt}\right) +
\frac 12 \frac{dm^2}{dt} G^2(t)\nonumber\\
&&+ \frac 14 \left(\lambda (t) + \frac{d\lambda}{dt} + 2\lambda \frac{d\ln G}{dt}\right)
G^4(t)\phi^2. \lb{63}\eea
From Eq.~(\ref{34}) and (\ref{46}) we have:
\be\frac{d\ln G}{dt} = - \frac 12 \gamma = \frac{3(e^2 + g^2)}{32\pi^2}.\lb{64}\ee
Using Eqs.~(\ref{47}), (\ref{48}), (\ref{62}) and (\ref{64}), it is easy to find the joint 
solution of equations
$V_{eff}(\phi_0^2) = V'_{eff}(\phi_0^2) = 0 $:
\be V'_{eff} (\phi_0^2) =\frac{1}{4} \left( - \lambda\beta_{(m^2)} +\lambda 
+ \beta_{\lambda} - \gamma\lambda\right)G^4(t_0)\phi_0^2 = 0,\lb{65}\ee
or
\be\beta_\lambda+\lambda\left(1-\gamma-\beta_{(m^2)}\right)=0. \lb{66}\ee
Eq.~(\ref{66}) gives the phase transition border, valid in any loop approximation. Putting into 
Eq.~(\ref{66}) the function $\beta_\lambda$, $\beta_{(m^2)}$ and $\gamma$, given by 
Eqs.~(\ref{44}-\ref{46}) and (\ref{56}) in the one--loop approximation, we obtain the following 
equation for the phase transition border:
\be{V'}_{eff}(\phi_0^2) = \left[\frac 3{16\pi^2}(e^2 + g^2)^2 + \lambda 
+ \frac 38 \frac{{\lambda }^2}{\pi^2}\right]G^4(t_0)\phi_0^2 = 0, \lb{67}\ee
or
\be\frac{3}{16\pi^2}(e^2+g^2)^2+\lambda+\frac{3}{8}\frac{\lambda^2}{\pi^2}=0.\lb{68}\ee
Using the Dirac relation (\ref{24}) and Eq.~(\ref{68}), we have:
\be g^4_{crit} + \frac{(2\pi)^4}{g^4_{crit}} + 8\pi^2 = 
- 2\lambda \left(\frac{8\pi^2}3 + \lambda \right). \lb{69}\ee
The curve (\ref{69}) is shown on the phase diagram $(\lambda ; g^2)$ of Fig.~\ref{f2} by the 
curve ``1", which describes a border between the ``Coulomb" phase with $V_{eff} \ge 0$ and the 
confinement ones, having $V_{eff}^{min} < 0$.

\vspace{1cm}\noindent{\bf 7.}
The next step is the calculation of the second derivative of the effective potential:
\bea{V''}_{eff}(\phi^2) &=& \frac {{V'}_{eff}(\phi^2)}{\phi^2} + \left( - \frac 12
 m^2+ \frac 12 \frac{d^2 m^2}{dt^2} + 2\frac{dm^2}{dt}\frac{d\ln G}{dt} +  
m^2\frac{d^2\ln G}{dt^2} \right.\nonumber\\
&& \left. + 2m^2\left(\frac{d\ln G}{dt}\right)^2\right)\frac {G^2}{\phi^2} + \left(
\frac 12 \frac{d\lambda }{dt} + \frac 14 \frac {d^2\lambda }
 {dt^2} + 2\frac{d\lambda }{dt}\frac{d\ln G}{dt}+ 2\lambda \frac{d\ln G}{dt}\right.\nonumber\\
&&  +\left. \lambda \frac{d^2\ln G}{dt^2} +
     4\lambda \left(\frac{d\ln G}{dt}\right)^2\right) G^4(t).  \lb{70}\eea
Let us consider now the case, when this second derivative changes its sign giving a maximum of 
$V_{eff}$, instead of the minimum at $\phi^2 = \phi_0^2$. Such a possibility is shown in 
Fig.~\ref{f1} by the curve ``2". Now the two additional minima at $\phi^2 = \phi_1^2$ and 
$\phi^2 = \phi_2^2$ appear in our theory. They correspond to the two different confinement 
phases related with the confinement of the electrically  charged particles. If these two minima 
are degenerate, then we have the following requirements:
\be V_{eff}(\phi_1^2) = V_{eff}(\phi_2^2) < 0    \lb{71}\ee
and
\be {V'}_{eff}(\phi_1^2) = {V'}_{eff}(\phi_2^2) = 0,   \lb{72}\ee
which describe the border between the confinement phases ``conf.~1" and ``conf.~2", presented 
in Fig.~\ref{f2}. This border is shown by the curve ``3" at the phase diagram $(\lambda; g^2)$ 
of Fig.~\ref{f2}. The curve ``3" meets the curve ``1" at the triple point $A$. It is obvious 
that, according to the illustration of Fig.~\ref{f1}, this triple point $A$ is given by the 
following requirements:
\be V_{eff}(\phi_0^2) = V'_{eff}(\phi_0^2) = V''_{eff}(\phi_0^2) = 0. \lb{73}\ee
In contrast to the requirements:
\be V_{eff}(\phi_0^2) = V'_{eff}(\phi_0^2) = 0, \lb{74}\ee
describing the curve ``1", we are going now to consider the joint solution of the following 
equations:
\be V_{eff}(\phi_0^2) = V''_{eff}(\phi_0^2) = 0 . \lb{75}\ee
It is possible to obtain this solution
using Eqs.~(\ref{70}), (\ref{62}), (\ref{47}), (\ref{48}) and (\ref{56}):
\be{\cal F}(\lambda , g^2) - \pi^2\lambda  + 2\pi^2(e^2 + g^2) = 0, \lb{76}\ee
where 
\bea{\cal F}(\lambda, g^2) &=& 5(e^2 + g^2)^3 +(24\pi^2 + 12\lambda)(e^2 + g^2)^2 - 
9\lambda^2(e^2 + g^2)\nonumber\\
&&+ 36\lambda^3 + 80\pi^2\lambda^2 + 64\pi^4\lambda . \lb{77}\eea
According to the Dirac relation, we have $e^2=(2\pi/g)^2$ and obtain
the curve ``3'' of Fig.~\ref{f2}, which represents the solution of Eq.~(\ref{76}) 
(together with (\ref{77})), which is equivalent to Eqs.~(\ref{75}).

The intersection of the curve ``1'' (or ``2'') by the curve ``3" gives the value of the triple 
point $A$ (or $B$), which corresponds to the critical values $\alpha_{crit}$ and 
$\tilde\alpha_{crit}$ in the Higgs model of scalar dyons.

The numerical calculations demonstrate that the triple point $A$ exists in the very 
neighbourhood of the maximum of the phase transition curve ``1'' of Fig.~\ref{f2}, and its 
position is given by the following values of $\lambda$ and $g^2$:
\be\lambda_A\approx -13.44 \qquad \mbox{and} \qquad g^2_A\approx 16.16. \lb{78}\ee
The triple point $B$ has its position near the minimum of the curve ``2'' and gives the
following values of $\lambda$ and $g^2$:
\be\lambda_B\approx -13.44 \qquad \mbox{and} \qquad g^2_B\approx 2.44. \lb{79}\ee
Here we see that the ratio:
\be\frac{e^2_A}{g^2_A}=\frac{4\pi^2}{g_A^4}\approx 0.15 \lb{80}\ee
describes 15\% contribution from $e^2_A$ to the value $g^2_A$. This result explains the
difference between the dyon model and scalar electrodynamics (see the review \ct{27}).

The negative values of $\lambda_{A,B}$ in the triple points are not dangerous, because they
are obtained for the renormalized quantity, and  point out that $\phi_0\ll M_{cut off}$,
what is quite possible.

It seems to us, that the triple point $A$ (or $B$) does not coincide with the maximum (minimum)
of the curve ``1'' (``2'') due to our one--loop approximation. We expect that they will 
coincide in the true theory.

Values (\ref{78}) and (\ref{79}) for $g^2_{A,B}$ correspond to the following critical fine 
structure constants:
\be\alpha_{crit}\approx 0.19, \qquad \mbox{and} \qquad \tilde\alpha_{crit}\approx 1.29,
\lb{81}\ee
which are in agreement with the result of lattice investigations (\ref{54}) for QED.

In the present theory of scalar dyons, the expressions (\ref{68}), (\ref{76}) and (\ref{77}) 
depend on the quantity $e^{*2}=e^2 + g^2$. This is a consequence of the dual symmetry 
(\ref{16}) and (\ref{17}). By dual transformations, we can reduce dyon fields to the electric 
fields with the effective electric charge $e^*$. From Eq.~(\ref{81}) the critical value of 
$\alpha^*=e^{*2}/4\pi$ is:
\be
           \alpha^*_{crit}\approx 1.48.          \lb{82}
\ee    
But the Dirac relation (\ref{24}), obtained in the item 2, distinguishes the electric and 
magnetic charges of dyons, when it interacts with the fields $A_{\mu}$ and $B_{\mu}$.
This is significant for the phenomenon of the string formation in QCD, considered 
in Ref.~\ct{5}. 

\vspace{1cm}\noindent{\bf Acknowledgements}

We deeply thank Prof.~H.B.~Nielsen for fruitful discussions and advises. One of the authors 
(L.V.L.) is indebted to the Institute of Mathematical Sciences (Chennai, India) and personally
Prof.~N.D.~Hari Dass for hospitality, financial support and interesting discussions.

This work was supported by the Russian Foundation for Basic Research (RFBR), project $N^o$
05--02--17642.

\clearpage\newpage
\bfi
\centering
\includegraphics[height=80mm,keepaspectratio=true]{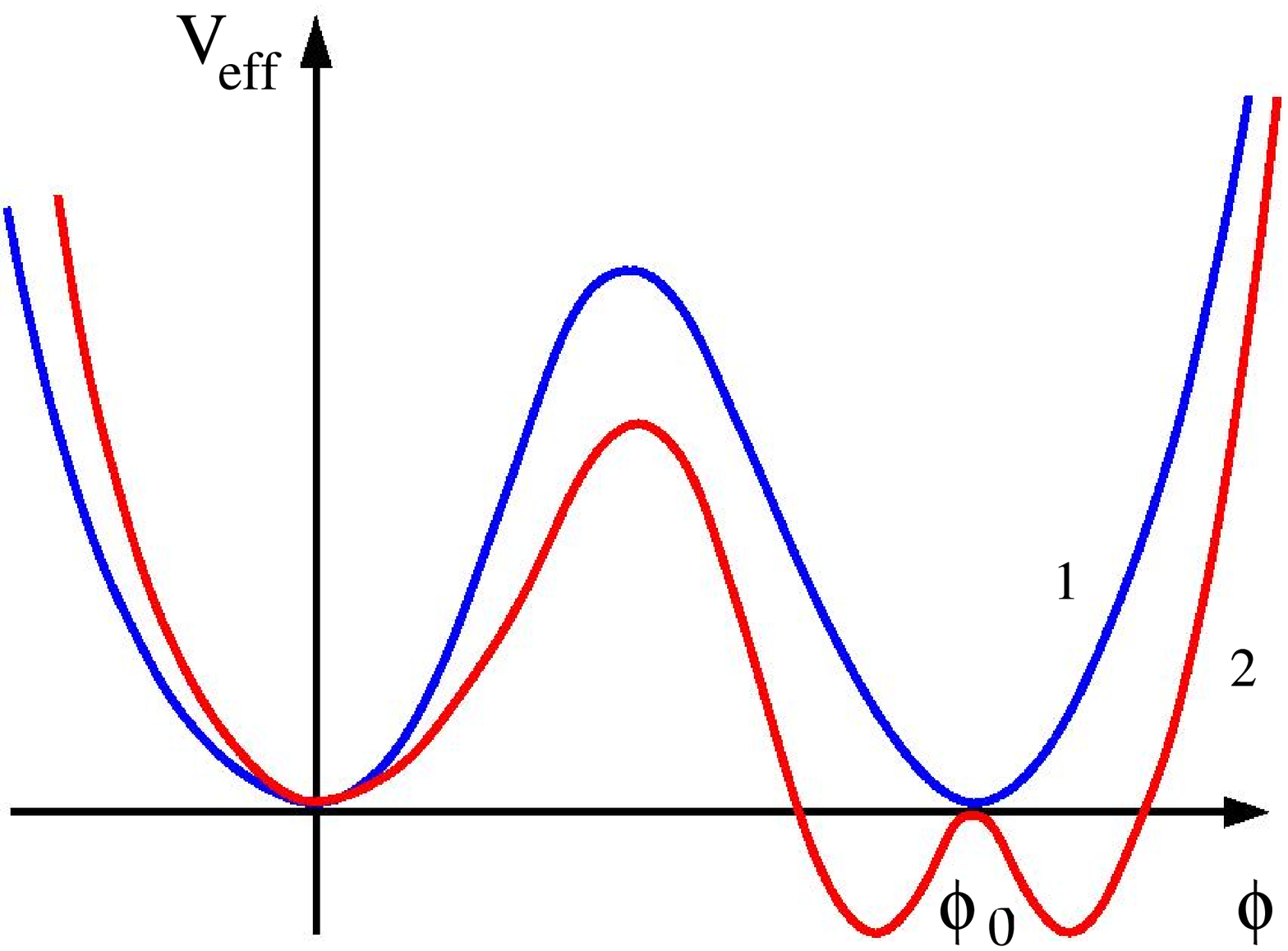}
\caption{The effective potential $V_{eff}$: the curve ``1'' corresponds to the 
``Coulomb--confinement'' phase transition; curve ``2'' describes the existence of two minima 
corresponding to the confinement phases.}
\lb{f1}
\efi

\clearpage\newpage
\bfi
\centering
\includegraphics[height=166mm,keepaspectratio=true,angle=-90]{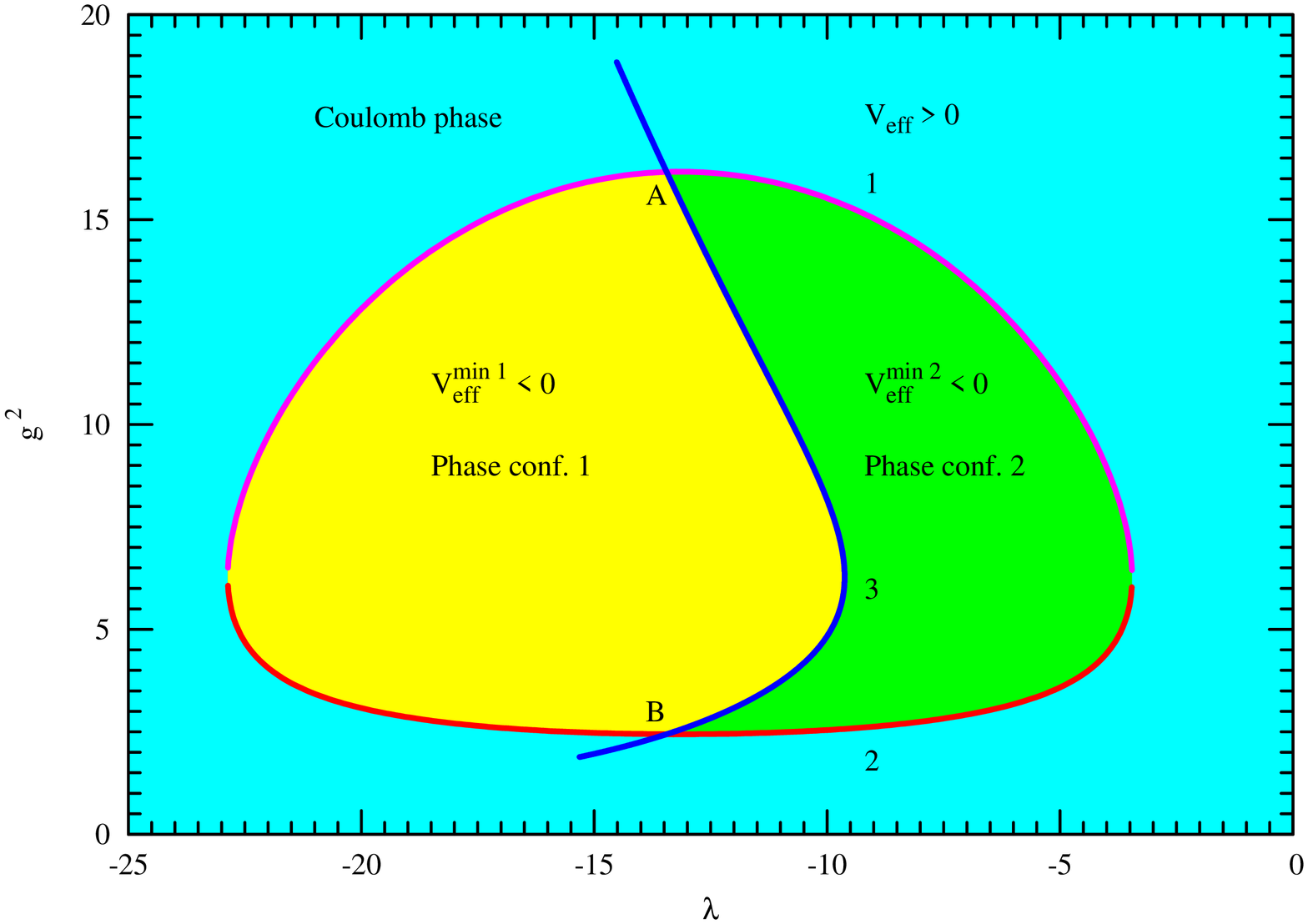}
\caption{The phase diagram $(\lambda;g^2)$ for the Higgs model of scalar abelian dyons. Curves 
``1'' and ``2'' separate ``Coulomb'' phase from confinement phase. Figure shows the existence 
of triple points $A(\lambda_A=-13.44;g_A^2=16.16)$ and $B(\lambda_B=-13.44;g_B^2=2.44)$. These 
triple points are boundary points of three phase transitions: the ``Coulomb'' phase and two 
confinement phases: ``conf.~1'' and ``conf.~2'', which are separated by the curve ``3''.}
\lb{f2}
\efi

\end{document}